\documentclass{cfm2011}
\usepackage[latin1]{inputenc}
\usepackage{graphicx}
\usepackage{color}

\title{Migration of anticyclonic vortices in the protoplanetary disk }
\author{C. Surville\inst{a}, P. Barge\inst{a}}

\instlabel{a}{Laboratoire d'Astrophysique de Marseille, Universit\'{e} de Provence, CNRS, 38 rue Fr\'{e}d\'{e}ric Joliot-Curie, F-13388 Marseille Cedex 13, France}

\begin{document}

\maketitle

\begin{abstract}
	\textit{ 
	This contribution describes the evolution of the protoplanetary disk using 2D numerical simulations. The 2D Euler equations are solved with the finite volume method. The numerical simulations are used to study the persistence and migration of anticyclonic vortices.  Two cases are presented: (1) vortices produced by the Rossby Waves Instability, (2) a non-linear vortex model initially implemented into the disk. The migration of the vortices is due to spiral density waves excited by the vortex in the gas of the disk.
	}
\end{abstract}


\section{Introduction}

        To understand protoplanetary disks' evolution is crucially important throughout the problematic of planetary formation. Therefore we are confronted with two major problems. The first appears to be a problem of time scale. Indeed the observations of protoplanetary disks and especially in the infrared excess resulting from dusts lead us to believe that the disk is dispersed after $10^7$ years (\cite{Briceno01}).        
       The second problem is the growth of  particles inside the disk. Actually, several theories have been proposed to explain how some micron sized particles present a the beginning of the disk's life can reach a kilometer size and form planetesimals or planetary cores (\cite{Weidenschilling77}, \cite{Goldreich79}, \cite{Barranco05}.  
        The principal boundary to the grain's growth is that the flow is sub-Keplerian whereas the motion of particles is keplerian.  When the size of solid particles is large in front of the mean free path of gas molecules, of the order of $10$ $cm$ to $1$ $m$, the particles undergo a wind which makes them fall onto the star in $100$ years typically. The coagulation or particle sedimentation's time is longer by at least one order of magnitude.
        
        Hence the particle's confinement in an environment where this radial migration is reduced, is favorable for the particle's growth within a time scale sufficiently short to create gaseous giants. Lots of studies bear on anticyclonic vortices because they capture particles in their core. (\cite{Barge95}, \cite{Heng10})
        Cyclones being quickly dispersed by the Keplerian shear, only anticyclonic vortices do survive longer enough in the flow. Several processes make it possible to form anticyclones, and particularly the Rossby Waves Instability or the baroclinic instability which lead to the formation of large anticyclonic structures.
        
        We propose to study those vortices in a 2D model, based on the Minimum Mass Solar Nebula Model which is the standard model of protoplanetary disks. 
In a first part, we will clarify the cylindrical Euler equations, and then a quasi linear approach of the Euler equations allows us to propose a vortex model.
In a second part, we will focus on different numerical simulations, with the study of several Rossby Wave Instability simulations and vortex models. 
To finish, we will explain those different simulation results, especially the vortices life time and their migration.

\section{The Euler equations and a vortex model}

	The standard model of 2D protoplanetary disks is based on the study of matter repartition in the Solar System and on disks observations. We suppose that density, temperature and pressure decrease like a power law of the radius. We assume that there is an hydrogen diatomic gas, polytropical with the following conditions in the thermodynamical equilibrium:
	   \begin{eqnarray*}  
	  \sigma_0 = 1700 \ \left( \frac{r}{1 AU} \right) ^p  \ g.cm^{-2} & &
	   T_0 = 280 \  \left( \frac{r}{1 AU} \right) ^q \ K \\
	  P_0 = k_B T_0 \ \sigma_0 & &
	   p + q = \gamma \ p
	   \end{eqnarray*} 
        with $ -2<p<0 $ and $ -1<q<0 $. We will choose here $ p = -3/2$, and $\gamma = 1.4$.
	\subsection{The Euler equations}
	The disk gas is in rotation in the Keplerian potential of the star of mass $M_{\ast}$. The Euler conservation equations make it possible to describe this gas evolution. In the cylindrical coordinates, we write down, with $\vec{V} = u \ \vec{u_r} + v \ \vec{u_{\theta}}$:
        \begin{eqnarray}
        &  & \partial_t \sigma + \frac{1}{ r} \partial_r ( r \sigma u) + \frac{1 }{ r}  \partial_\theta (\sigma v) = 0 \\
        &  & \partial_t \sigma u + \frac{1}{ r} \partial_r ( r \sigma u^2) + \frac{1 }{ r}  \partial_\theta (\sigma u v) = \sigma \frac{v^2}{r} - \sigma \frac{v_k^2}{r} - \partial_r P  \\
        &  & \partial_t \sigma v + \frac{1}{ r} \partial_r ( r \sigma v u) + \frac{1 }{ r}  \partial_\theta (\sigma v^2) = - \sigma \frac{u v}{r}  - \frac{1 }{ r}  \partial_\theta P \\
        &  & \partial_t \sigma e  + \frac{1}{ r} \partial_r ( r (\sigma e + P) u ) + \frac{1 }{ r}  \partial_\theta ((\sigma e + P)  v) = - \sigma u \frac{v_k^2}{r}  
        \end{eqnarray}
         with $ v_k = \sqrt{\frac{G M_{\ast}}{r}}$, and the total gas energy $\sigma e = \frac{P}{\gamma - 1} + \frac{\sigma}{2} ( u^2 + v^2)$.         
                 
        A stationary solution associated with thermodynamical quantities $\sigma_0$ and $P_0$ is given by the following velocity $\vec{V_0} = u_0 \ \vec{u_r} + v_0 \ \vec{u_{\theta}}$:
        \begin{eqnarray}
        & & u_0 = 0 \\
        & & v_0 = \left( v_k^2 + \gamma p \ \frac{P_0}{\sigma_0} \right) ^{1/2}
        \end{eqnarray}
        where the second term of the right member derives from $\frac{r}{\sigma_0} \partial_r P_0$. We can notice that the gas is in sub-Keplerian rotation because of its pressure. We will call it the \emph{stationary solution}.
        \subsection{A vortex model}
        We search a quasi stationary non-axisymetrical solution of the Euler equations for this flow.
We set down:
        \begin{eqnarray}
        & \vec{V} = \vec{V_0} + \vec{\tilde{V}} \\
        & \vec{\tilde{V}} = \tilde{u} \ \vec{u_r} + \tilde{v} \ \vec{u_{\theta}} \\
        & \sigma = \sigma_0 \ \tilde{\sigma} \\
        & P = P_0 \ \tilde{P} \\
        & c_s^2 = \frac{P}{\sigma} = c_{s_0}^2 \ \frac{\tilde{P}}{\tilde{\sigma}}
        \end{eqnarray}        
        If we get into a rotating frame with $\vec{V_0}$, the continuity and velocity equations become:
        \begin{eqnarray}
        & & \partial_t \sigma + \tilde{u}  \partial_r \sigma + \frac{\tilde{v}}{r} \partial_{\theta} \sigma + \sigma \left[ \frac{1}{r} \partial_r (r \tilde{u}) + \frac{1}{r} \partial_{\theta} \tilde{v} \right] = 0 \\
        & & \partial_t \tilde{u} + \tilde{u}  \partial_r \tilde{u} + \frac{\tilde{v}}{r} \partial_{\theta} \tilde{u} = \tilde{v} \frac{\tilde{v}}{r} + \tilde{v} \frac{2 v_0}{r} - c_{s_0}^2 
        \left[ \frac{\gamma p}{r} \left( \frac{\tilde{P}}{\tilde{\sigma}} - 1 \right) + \frac{1}{\tilde{\sigma}} \partial_r \tilde{P} \right] \\
	  & & \partial_t \tilde{v} + \tilde{u}  \partial_r \tilde{v} + \frac{\tilde{v}}{r} \partial_{\theta} \tilde{v} = -\tilde{u} \frac{\tilde{v}}{r} - \tilde{u} \frac{v_0 + r \partial_r v_0}{r} 
	  - \frac{c_{s_0}^2}{\tilde{\sigma}} \frac{1}{r} \partial_{\theta} \tilde{P}
        \end{eqnarray}
        We search a stationary solution, so we lay $\partial_t = 0$. If we suppose the terms in $\tilde{u}\tilde{v}$, $\tilde{v}^2$, $\tilde{u}  \partial_r$ and $\frac{\tilde{v}}{r} \partial_{\theta}$ to be small in front of the other terms, then the velocity field is given by:
	  \begin{eqnarray}
	  & & \tilde{u} = - v_0 \ \frac{M_{a_0}^{-2}}{1+r \partial_r ln(v_0)} \ \frac{r}{\tilde{\sigma}} \ \frac{1}{r} \partial_{\theta} \tilde{P} \\
	  & & \tilde{v} = v_0 \ \frac{M_{a_0}^{-2}}{2} \left[ \gamma p \left( \frac{\tilde{P}}{\tilde{\sigma}} - 1 \right) + \frac{r}{\tilde{\sigma}} \ \partial_r \tilde{P} \right] 
	  \end{eqnarray}
	  where we define the Mach number of the stationary solution $M_{a_0}^2 = \frac{v_0^2}{c_{s_0}^2} \sim 10^3 \left(\frac{r}{1 AU}\right) ^ {-(q+1)}$.
	  
	  Thus, considering a polytropical gas, the only prescription of $\tilde{\sigma}(r, \theta)$ gives a quasi stationary solution. Our vortex model is given by:
	  \begin{eqnarray}
	  & & \tilde{\sigma} = 1 + A \ exp \left[ - \frac{r_0^2}{\omega_r^2} \left[ \left(\frac{r}{r_0} - 1\right)^2 + \frac{1}{\chi^2} \left(\theta - \theta_0\right)^2 \right] \right]
        \end{eqnarray}
        where $r_0$ and $\theta_0$ define the vortex position, $A$ the amplitude at the vortex center, $\omega_r$ the radial width and $\chi$ the aspect ratio of the vortex.

\section{Numerical simulations}
	  The temporal solving of the Euler equations is made with a numerical scheme using the finite volume method, an exact Riemman solver, and in which the calculation method of the radial flux was improved regarding the classical MUSCL scheme. The time integration is based on the 2d order  Runge-Kutta method. The temporal stability is verified which allows to integrate the solution for more than 1000 disk rotations without significant errors ($10^{-8}$ on the speeds and $10^{-6}$ on the density). The boundary conditions are the continuity to the stationary solution $\sigma_0$, $P_0$, $\vec{V_0}$ in the radial direction, and periodic conditions in the azimuthal direction.
	  
	  We use a polar grid, with $n_r=500$ cells in the radial directio and $n_{\theta}=1000$ cells in the azimuthal direction. Moreover four ghost cells are used outside the domain to prevent any perturbation due to the boundaries. Finally, the disk extends from $5 \ AU$ to $10\ AU$.
\subsection{The Rossby Waves Instability}
	The first set of simulations is the study of the so-called Rossby Waves Instability. We will not look closely into it so we refer to the articles \cite{Li00} and \cite{Lovelace99}. Those studies emphasize the fact that an axisymetrical perturbation with a Gaussian radial profile can trigger it if the height and the width of this perturbation are sufficient. Hence we introduce the following initial conditions in our simulations:
	\begin{eqnarray}
	& & \tilde{\sigma}_{ini} = 1 + f \\
	& & \tilde{P}_{ini} = (1 + f)^{\gamma} \\
	& & \tilde{u}_{ini} = 0 \\
	& & \tilde{v}_{ini} = v_0 \ M_{a_0}^{-2} \gamma (1+f)^{\gamma-1} \left[ p \left( 1 - (1+f)^{1-\gamma} \right) + \frac{r}{1+f} \ \partial_r f \right]
	\end{eqnarray}
	where $f$ is the gaussian perturbation. We set $f = f_0 \ exp \left[ -\frac{(r-r_0)^2}{\omega_f^2} \right]$ and $r_0=7.5\ AU$.
	We have performed $6$ simulations whith the following parameters $f_0$ and $\omega_f$:
	\begin{center}
	\begin{tabular}{|c||c|c|c|c|c|c|}
	\hline
	 $f_0$ & $20\%$ & $20\%$ & $20\%$ & $30\%$ & $30\%$ & $30\%$ \\
	 \hline
	 $\omega_f$ & $0.2 \ AU$ & $0.3 \ AU$ & $0.4 \ AU$ & $0.2 \ AU$ & $0.3 \ AU$ & $0.4 \ AU$ \\
	 \hline
	 \end{tabular}
	 \end{center}
	 
	\begin{figure}
	\begin{center}
	\begin{tabular}{ccc}
       \includegraphics[width=5cm]{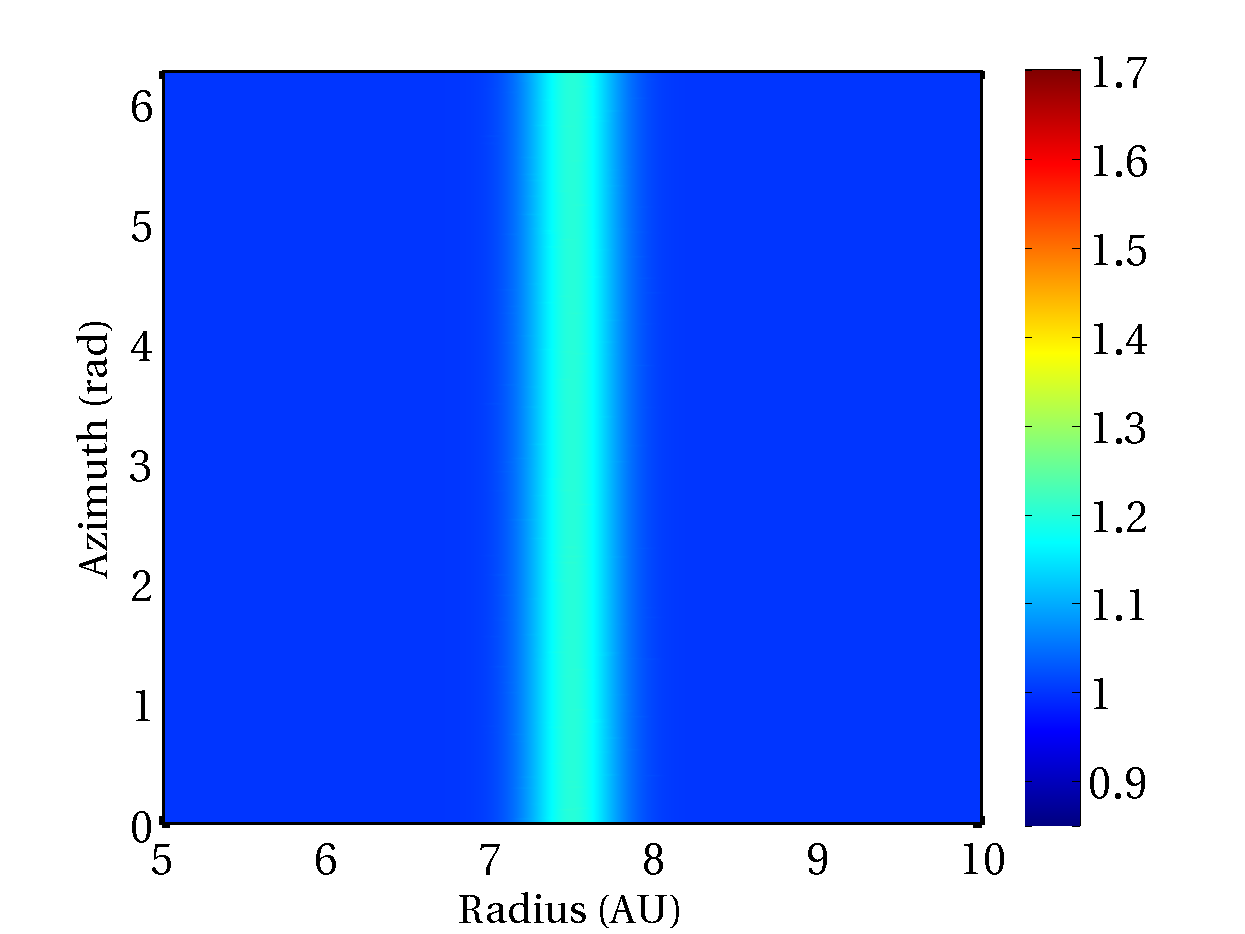} &
       \includegraphics[width=5cm]{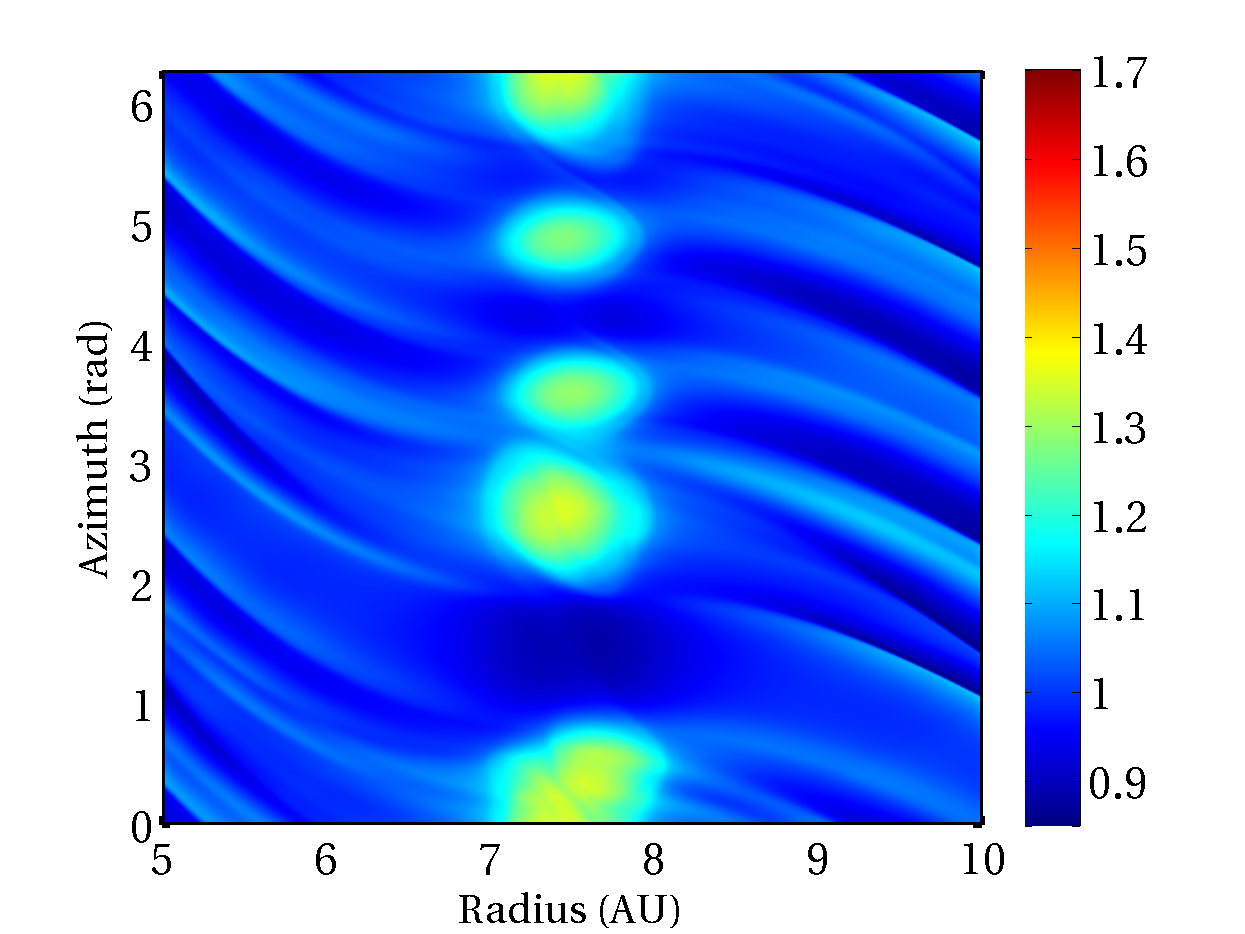} &
       \includegraphics[width=5cm] {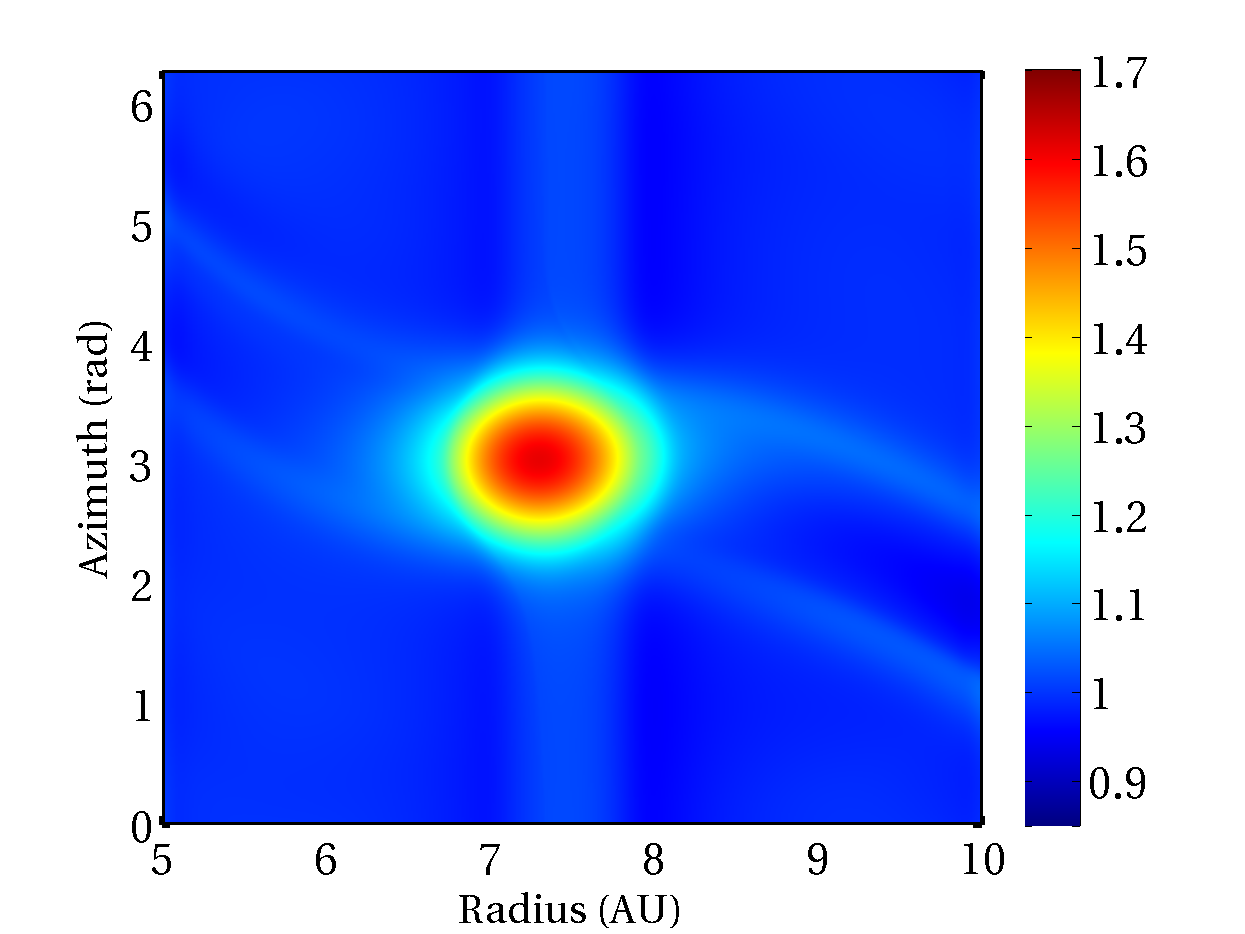} \\
       (a1) & (a2) & (a3) \\
       \includegraphics[width=5cm]{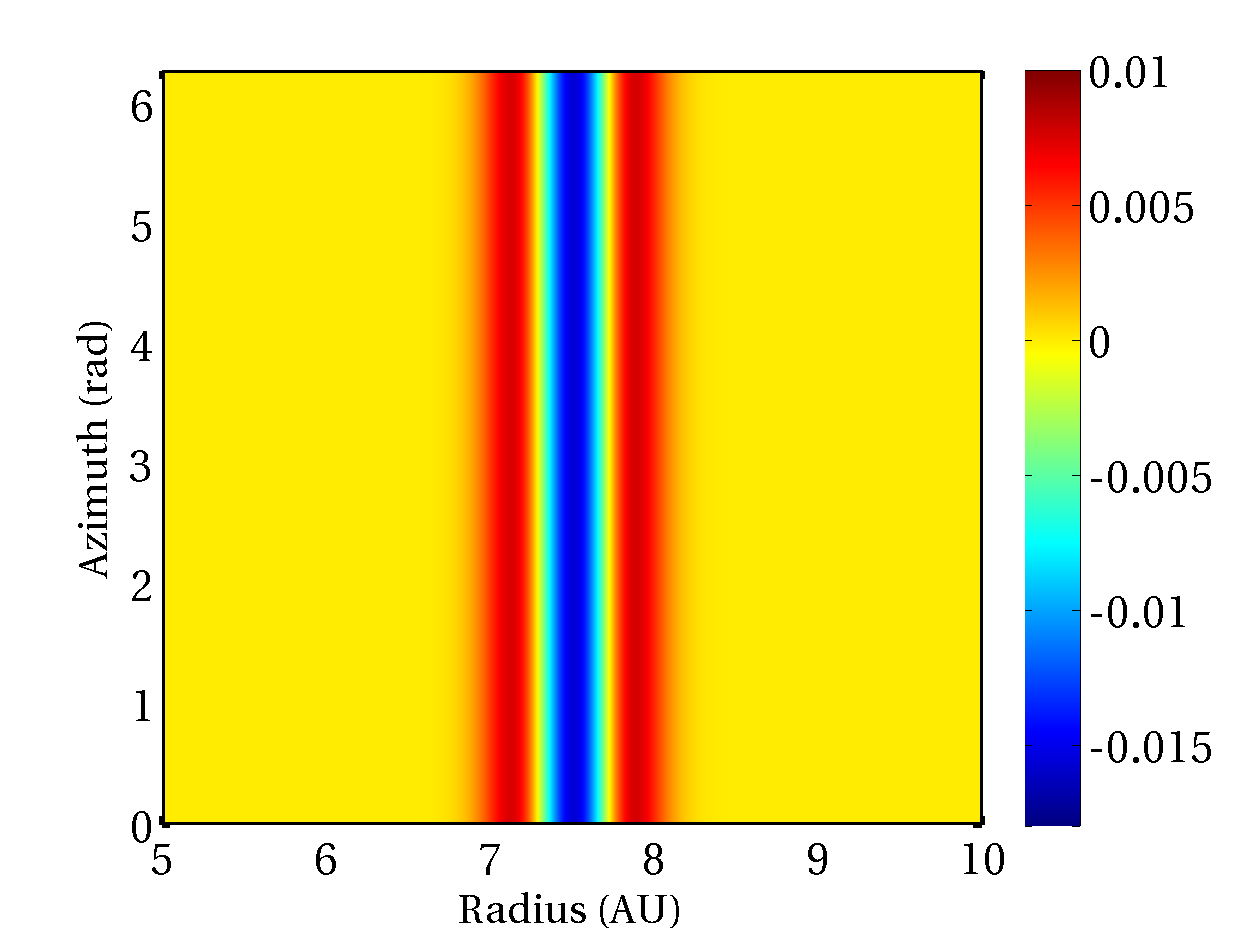} &
       \includegraphics[width=5cm]{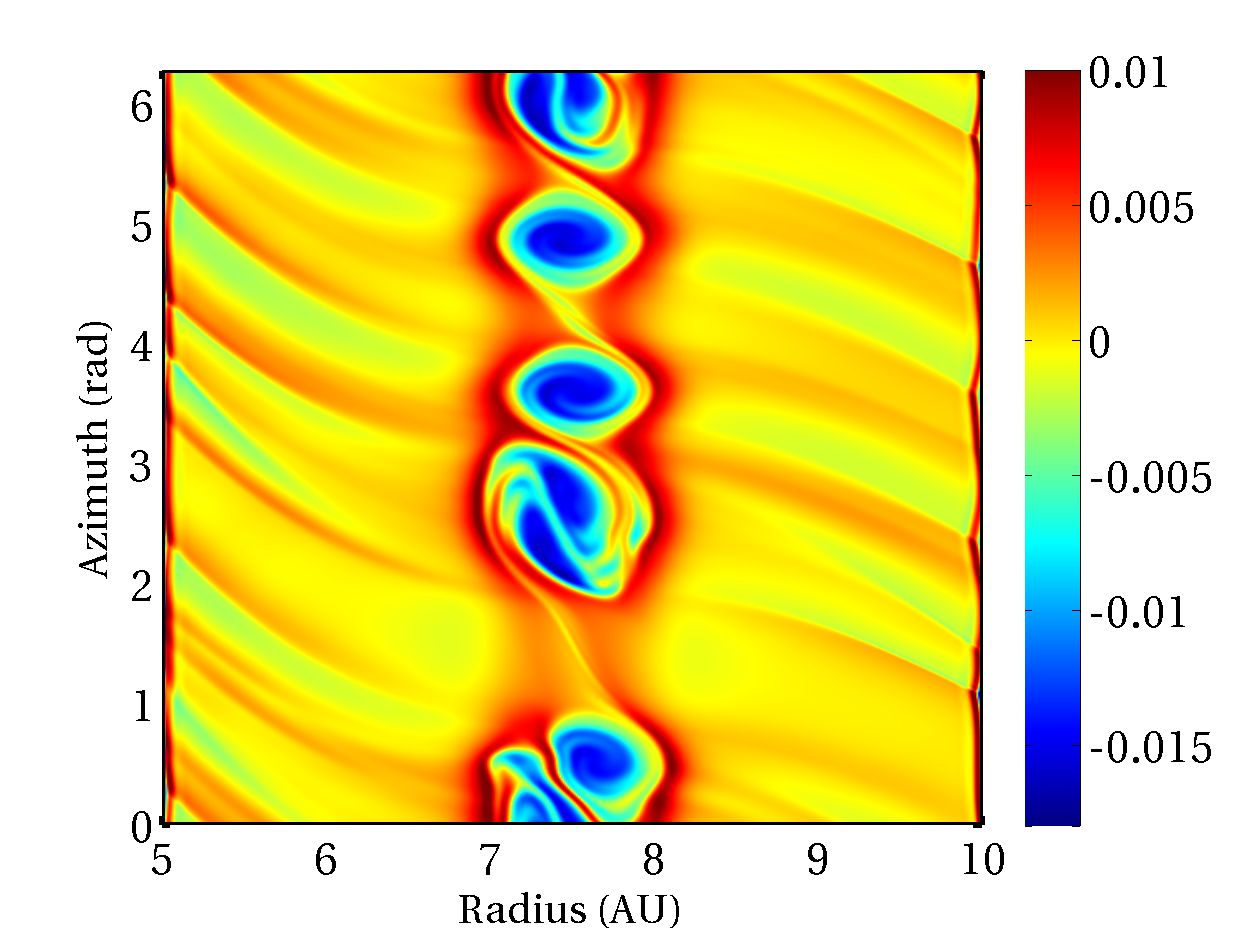} &
       \includegraphics[width=5cm] {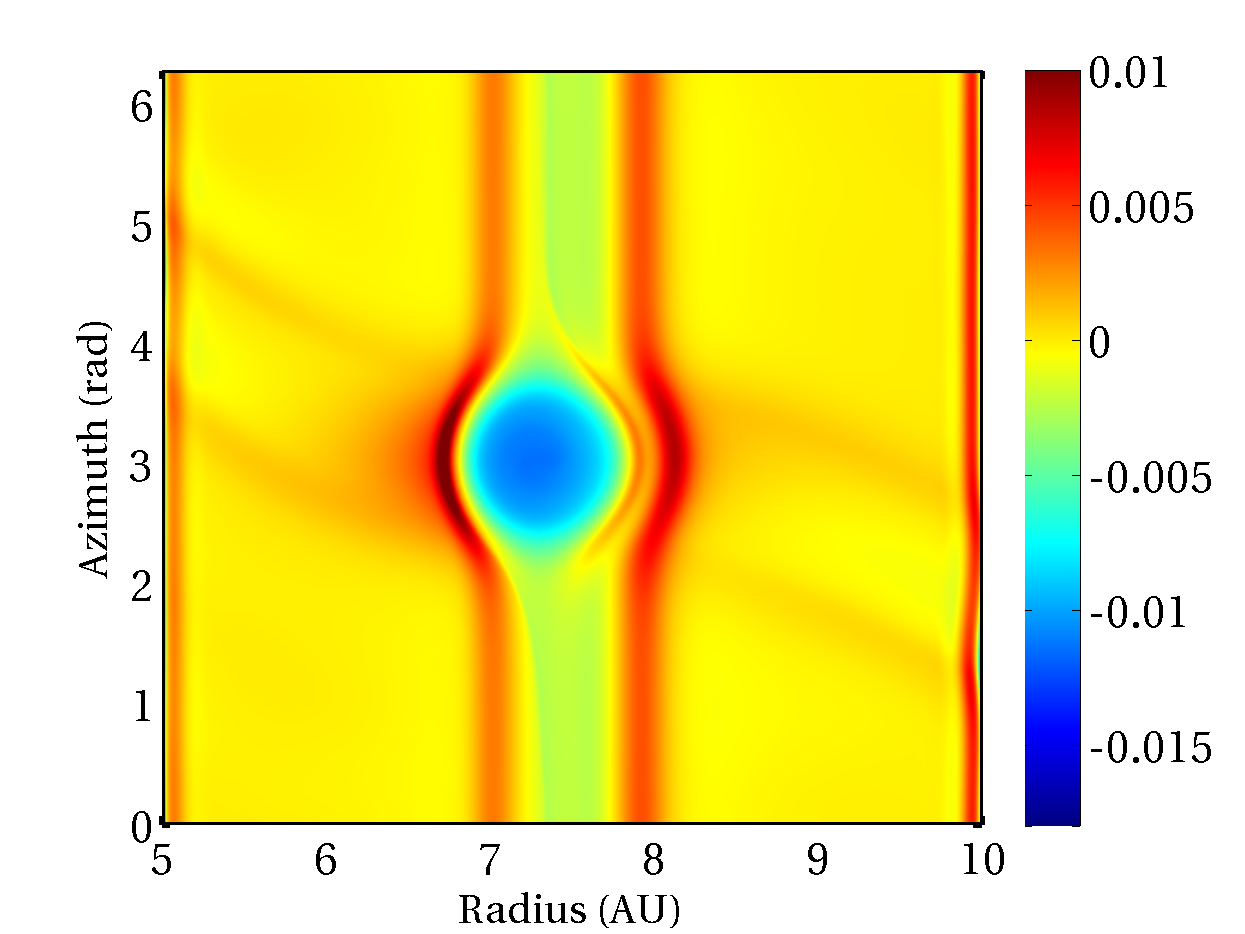}\\
       (b1) & (b2) & (b3) \\
      \end{tabular}
                    \caption{\label{Rossby_Insta} Density and vorticity relatively to the stationary solution at the initial time (a1 et b1), after $16$ rotations (a2 et b2) and after $180$ rotations (a3 et b3) for the parameters $f_0 = 20 \%$ and $\omega_f = 0.3 \ AU$}
       \end{center}
      \end{figure}

\subsection{Vortex models}
	 We have studied several vortex models with different parameters. The vortex position is $r_0 = 7.5 \ AU$ et $\theta_0 = 3 \ rad$. Three types of simulations have been performed:
	 \begin{center}
	 \begin{tabular}{|c||c||c||c|}
	 \hline
	 $\omega_r \ (AU)$ & $0.6 $ & $0.6$ &
	 \begin{tabular}{c|c|c}
	 $0.5$ & $0.7$ & $0.9$ 
	 \end{tabular}
	 \\
	 \hline
	 $\chi $ & $10$ & 
	 \begin{tabular}{c|c|c|c}
	 $12.5$ & $10$ & $7.5$ & $5$ 
	 \end{tabular} &
	 \begin{tabular}{c|c|c}
	 $12$ & $8.57$ & $6.67$
	 \end{tabular}
	 \\
	 \hline
	 $A$ & 
	 \begin{tabular}{c|c|c|c|c}
	 $0.5$ & $0.6$ & $0.7$ & $0.8$ & $0.9$
	 \end{tabular}
	 & $0.6$ & $0.6$
	 \\
	 \hline
	 \end{tabular}	 
	 \end{center}
	 
	\begin{figure}
	\begin{center}
	\begin{tabular}{cc}
       \includegraphics[width=5cm]{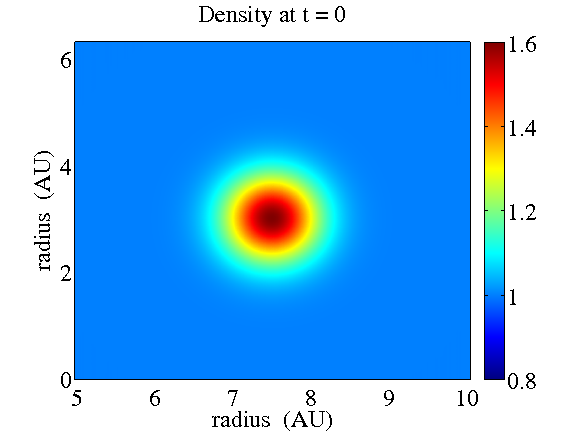} &
       \includegraphics[width=5cm] {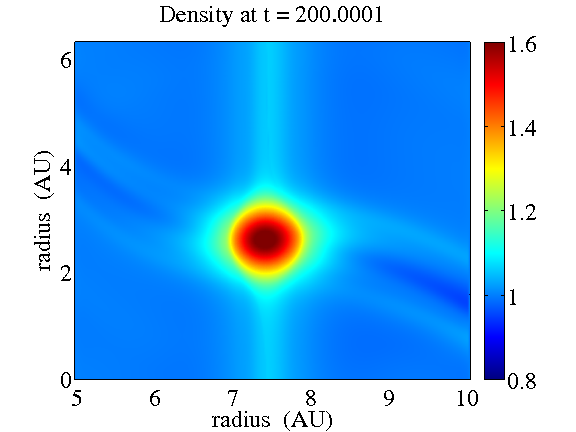} \\
       (a1) & (a2) \\
       \includegraphics[width=5cm]{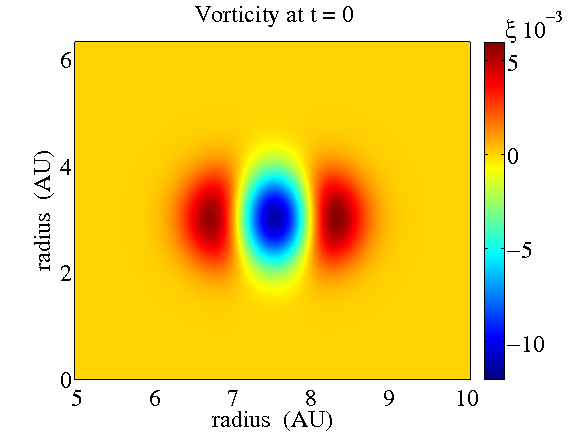} &
       \includegraphics[width=5cm] {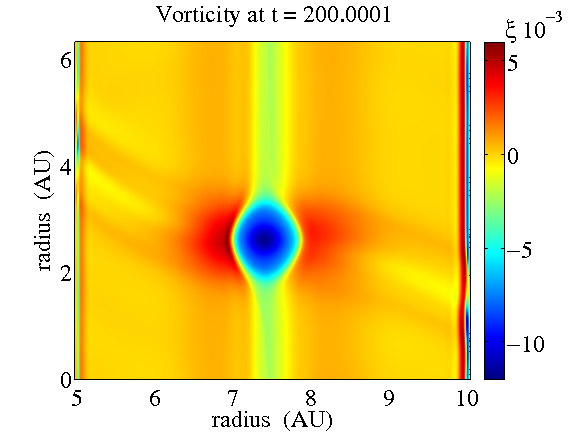} \\
       (b1) & (b2) \\      
      \end{tabular}
                    \caption{\label{Vortex_evo} Density and vorticity relatively to the stationary solution at the initial time (a1 et b1) and after $200$ rotations (a2 et b2) for a model with the parameters $A = 60 \%$, $\omega_r = 0.6 \ AU$ and $\chi = 10$}
       \end{center}
      \end{figure}

\section{Discussion}
\begin{figure}
	\begin{center}
	\begin{tabular}{cc}
       \includegraphics[width=6cm]{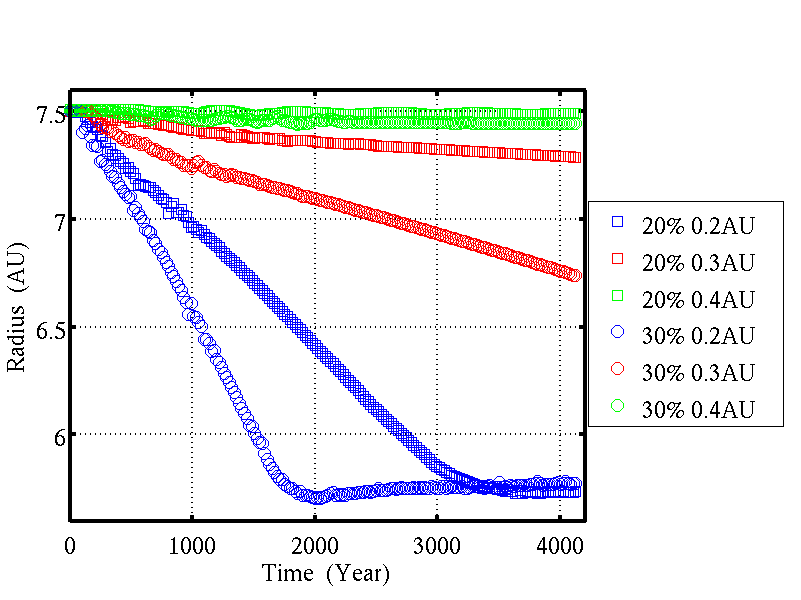} &
      \includegraphics[width=6cm] {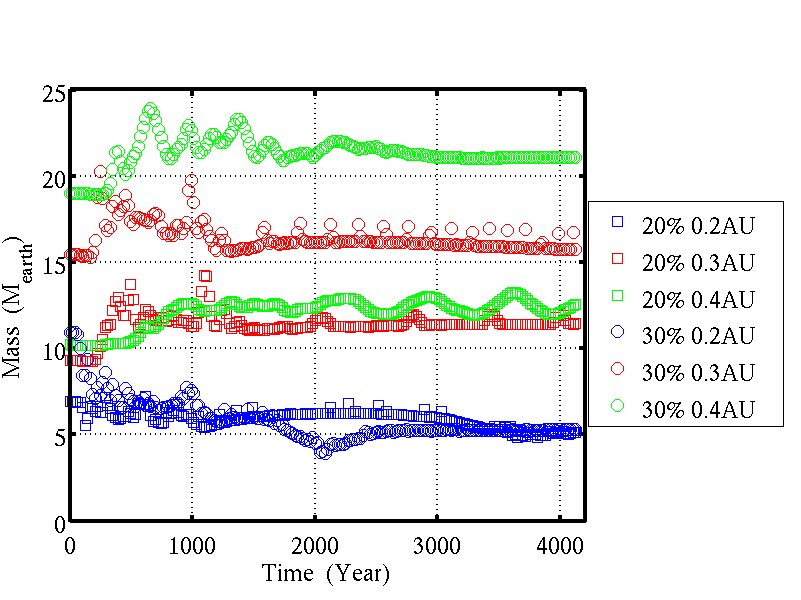} \\
      (a) & (b) \\
      \end{tabular}
       \caption{\label{Position_Mass} Time evolution of the vortex position (a) and the vortex mass (b)  for different simulations}
       \end{center}
      \end{figure}

\begin{figure}
	\begin{center}
	\begin{tabular}{cc}
       \includegraphics[width=5cm]{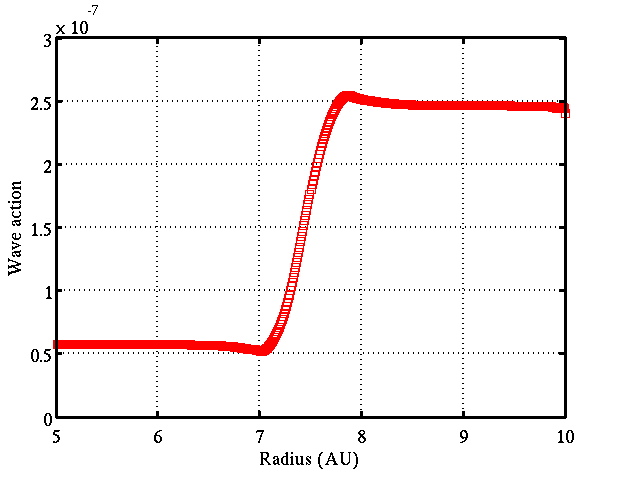} &
      \includegraphics[width=5cm] {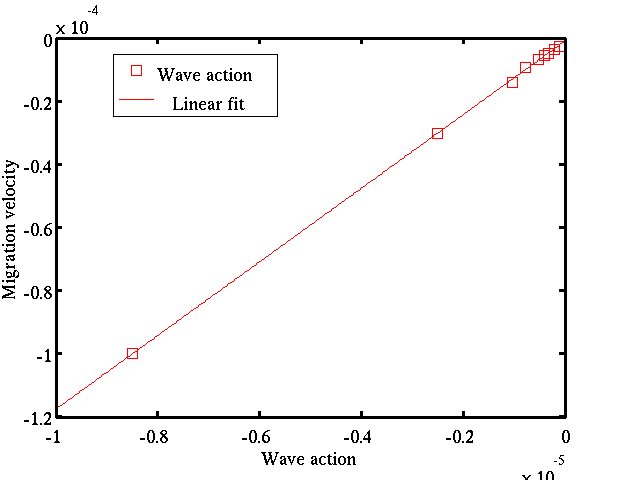} \\
      (a) & (b) \\
      \end{tabular}
       \caption{\label{Radial_Flux} Radial transport of kinnetic momentum (a) and its variation with the migration velocity (b) for different simulations}
       \end{center}
      \end{figure}

	The evolution of vortices is similar either they are issued from the Rossby Waves Instability or imposed in the disk. First, the several simulations confirm that vortices are robust solutions of the Euler equations. The Figures \ref{Rossby_Insta} and \ref{Vortex_evo}, which show the time evolution of different simulations, highlight that these structures conserve their mass and their vorticity for more than a hundreds rotations. They continue to exist during more than $10^4$ years, which represents an important fraction of the disk life. As a consequence, it can not be considered as transitory solutions.
	\begin{figure}
	\begin{center}
	\begin{tabular}{cc}
       \includegraphics[width=5cm]{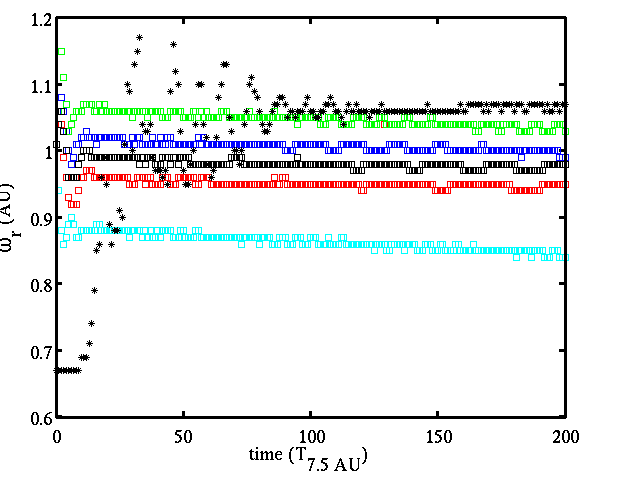} &
      \includegraphics[width=5cm] {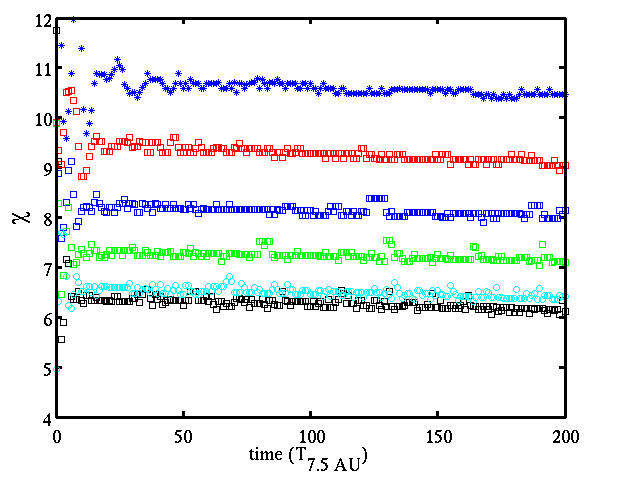} \\
      (a) & (b) \\
      \end{tabular}
       \caption{\label{Khi_r_Khi_theta} Time evolution of $\omega_r$ (a) and $\chi$ (b) for different simulations}
       \end{center}
      \end{figure}
	The range of solutions is vaste in terms of radial width and of aspect ratios. The  Figure \ref{Khi_r_Khi_theta} shows these different parameters concerning different simulations. Only the radial width is  more constrained. Indeed the  shear of the surrounding gas restricts the radial width to the disk scale height:
	\begin{eqnarray}
	& & H = \frac{r c_{s_0}}{v_0} \sim 0.45 \textrm{ \textit{AU} à } 7.5 \textrm{  \textit{AU}}
	\end{eqnarray}
	Nevertheless, most of the vortices have a radial width $\omega_r$ wider than $2H$ and are stable (\cite{Lesur09}). It seems that the flow is modified when the vortex is present, thus the structure undergoes a weak shearing between its inner and outer boundaries. 
The most noticeable phenomenon is the radial migration of vortices. They loose their kinetic momentum with a constant rate as the Figure \ref{Position_Mass} illustrates. The emission of density waves whose spiral pattern is due to the Keplerian shear, is systematic and responsible for the kinetic momentum transport. The article \cite{Paardekooper10} proposes a linear study of the waves emitted by a perturbation and introduces the term of radial flux of the kinetic momentum which is written like the following equation:
	\begin{eqnarray}
	& & F_r = 2 \pi r < r \sigma \tilde{v} \tilde{u} >
	\end{eqnarray}
      where $< \  >$ stands for the azimuthal average.  
Such a flux is not linked with linear waves. Nevertheless, we emphasize that the waves of our simulated vortices have radial transport which is constant in time. The wave of the inner part has an inferior rate than the one of the outer wave (Figure \ref{Radial_Flux}.a). The vortex situated at the center of the gap loses kinetic momentum. This radial flux is directed to the outside, as a consequence the disk and the vortex migrate onto the star. We illustrate Figure \ref{Radial_Flux}.b that this transport inside the waves is proportional to the vortex migration speed.
        
      The vortices migration speed is situated between $0.2 - 2 \ \frac{AU}{10^3 \ ans}$. This migration appears thus to be fundamental in planet formation models and disk evolution scenarios. This migration can justify hot jupiters formation or a model in which the accretion is kept going by a huge number of vortices split in the disk. The vortex model that we have developed allows us to better understand this non-linear phenomenon and we plan to present the continuation of our work in a future publication.


\end{document}